\documentclass[manuscript]{acmart}

\AtBeginDocument{%
  \providecommand\BibTeX{{%
    \normalfont B\kern-0.5em{\scshape i\kern-0.25em b}\kern-0.8em\TeX}}}

\setcopyright{acmlicensed}
\copyrightyear{2024}
\acmYear{2024}
\acmDOI{XXXXXXX.XXXXXXX}

\acmConference[Conference acronym 'XX]{Make sure to enter the correct conference title from your rights confirmation emai}{June 03--05, 2018}{Woodstock, NY}
\acmISBN{978-1-4503-XXXX-X/18/06}

\begin{document}

\title[Exploring the Design of Virtual Reality Museums to Support Remote Visitation With Older Adults]{Exploring the Design of Virtual Reality Museums to Support Remote Visitation With Older Adults
}

\author{Jingling Zhang}
\authornote{Both authors contributed equally to this research.}
\affiliation{%
  \institution{The Hong Kong University of Science and Technology (Guangzhou)}
  \city{Guangzhou}
  \country{China}}
\email{jzhang898@connect.hkust-gz.edu.cn}
\orcid{0009-0006-6555-5201}

\author{Qianjie Wei}
\authornotemark[1]
\affiliation{%
  \institution{The Hong Kong University of Science and Technology (Guangzhou)}
  \city{Guangzhou}
  \country{China}}
\email{qwei883@connect.hkust-gz.edu.cn}
\orcid{0009-0001-4429-4499}

\author{Xiaoying Wei}
\affiliation{%
  \institution{The Hong Kong University of Science and Technology}
  \city{Hong Kong SAR}
  \country{China}}
\email{xweias@connect.ust.hk}
\orcid{0000-0003-3837-2638}

\author{Mingming Fan}
\authornote{Corresponding author}
\affiliation{%
  \institution{The Hong Kong University of Science and Technology (Guangzhou)}
  \city{Guangzhou}
  \country{China}}
\affiliation{%
  \institution{The Hong Kong University of Science and Technology}
  \city{Hong Kong SAR}
  \country{China}}
\email{mingmingfan@ust.hk}
\orcid{0000-0002-0356-4712}

\renewcommand{\shortauthors}{Zhang, et al.}


\begin{abstract}

Virtual Reality (VR) museums provide immersive visiting experiences. Despite growing efforts in VR museum design optimization, limited research addresses its efficacy for older adults. We sought to investigate the challenges of and preferences for VR museum visits among older adults through a user-centered participatory workshop. Our preliminary findings illuminate issues regarding spatial navigation, interpretive descriptions, collective aspiration for augmented multi-sensory interactions, and imagined content visualization. Based on our preliminary findings, we discuss potential design principles for enhancing the accessibility of VR museums for older adults.

\end{abstract}

\ccsdesc[500]{Human-centered computing~Empirical studies in HCI; Empirical studies in accessibility}

\keywords{VR, virtual reality, museum, aging, older adults}




\maketitle

\section{INTRODUCTION}

Museums play a pivotal role in the lives of older adults. Prior research underscores the manifold benefits of museum visits, including the facilitation of family cohesion \cite{blud1990social, zhou2019benefits, jaszberenyi2021family, temple2023habits}, promotion of lifelong learning \cite{hsieh2010museum}, mitigation of social isolation \cite{hsieh2010museum, todd2017museum, camic2017museums}, and potential deterrence of cognitive decline such as dementia \cite{fancourt2018cultural}. Nonetheless, older adults frequently face impediments associated with physical mobility when visiting physical museums \cite{temple2023habits, pisoni2020mediating}. This is attributed to mobility limitations and long walking distances within museum premises. To mitigate these issues, older adults access the web-based versions of virtual museums through devices such as computers or mobile phones \cite{vasudavan2015preliminary, lafontaine2022virtual}. Although these virtual museums allow people to access them remotely\cite{pisoni2020mediating, hilton2019don}, older adults found them lacked the immersive qualities necessary for a comprehensive understanding of the museum's ambiance, grandeur, and the true dimensions of exhibited artworks \cite{lafontaine2022virtual}.

VR presents a promising avenue to surmount these constraints by offering immersive experiences that engender a profound sense of physical presence \cite{pisoni2019interactive} and foster heightened engagement between remote visitors and the artworks \cite{pisoni2020mediating}. Although the burgeoning research efforts aimed at optimizing the design of VR museums \cite{10.1145/3369394, 10.1145/3615522.3615544, 10.1145/3013971.3014018, 10.1145/3609987.3610006}, there remains a dearth of studies exploring its efficacy specifically concerning older adults. Older adults may encounter unique challenges and preferences when visiting VR museums. Previous research indicated that there is a risk of older adults becoming disoriented in VR environments \cite{castilla2013process}. Additionally, older individuals exhibit unique cognitive abilities and styles compared to younger groups. For instance, cognitive styles such as ENTP (Extroversion, Intuition, Thinking, Perceiving) are more prevalent in younger populations, while styles like INTJ (Introversion, Intuition, Thinking, Judging) are more commonly observed in older adults~\cite{antoniou2008reflections, antoniou2010modeling}. Therefore, it is imperative to design VR museum experiences tailored to the unique challenges and preferences of older individuals. This approach aims to improve the accessibility and appeal of VR museums for this demographic, ensuring they could fully engage with and benefit from these virtual environments. As such, we propose two Research Questions (RQs):

\begin{itemize}
\item \textbf{RQ1:} What are the challenges of and preferences for visiting VR museums among older adults? 
\item \textbf{RQ2:} How could VR museums be designed to be more accessible for older adults?
\end{itemize}


To answer the two RQs, we conducted a user-centered participatory workshop with 6 older adults. Our preliminary findings revealed some challenges when older adults visit VR museums: 1) Participants had difficulty in spatial positioning and navigation within the VR museum environment; 2) Participants universally perceived the interpretive descriptions provided by the system as not matching their cognitive capacities. Our preliminary findings also concluded some preferences of older adults for the VR museum: 1) There was a collective desire among participants for heightened multi-sensory interactions; 2) Participants showed great interest in visualizing what they imagined in their minds to facilitate communication with others; 3) Participants exhibited varied preferences regarding exploration modes. We further identified the potential design implications and outlined pertinent avenues for future research to better enhance the accessibility of VR museums for older individuals.

In summary, we make the following contributions: 1) we made an initial investigation into the challenges and preferences of visiting VR museums for older adults, and 2) we proposed initial design principles for enhancing the accessibility of VR museums for older adults.


\section{BACKGROUND}
\subsection{Virtual Museums Visitation for Older Adults}

Museums play a pivotal role in shaping the cultural milieu of older individuals. Extant literature has evidenced the manifold advantages of museum visits, including the augmentation of familial bonds \cite{blud1990social, zhou2019benefits, jaszberenyi2021family, temple2023habits}, facilitation of lifelong learning \cite{hsieh2010museum}, mitigation of social isolation \cite{hsieh2010museum, todd2017museum, camic2017museums}, and potential deterrent effects against dementia \cite{fancourt2018cultural}. Nonetheless, due to the physical challenges older adults face, such as mobility issues, their active involvement, and participation in traditional museums are hindered \cite{temple2023habits, pisoni2020mediating}.

The utilization of virtual museums, enabling remote access, stands poised to mitigate the mobility constraints prevalent among older adults \cite{pisoni2020mediating, hilton2019don}. However, presently, older individuals predominantly access the web-based iterations of virtual museums via devices such as computers or mobile phones \cite{vasudavan2015preliminary, lafontaine2022virtual}. For example, previous research presented a series of virtual museum visits through computers during the COVID-19 pandemic \cite{lafontaine2022virtual}. However, due to the lack of an immersive experience for older adults when interacting with virtual museums through web interfaces, they miss out on a sense of the overall museum ambiance and face difficulties in grasping the scale of paintings \cite{lafontaine2022virtual}. Recent research suggests that VR holds the potential to overcome these limitations by facilitating immersive experiences \cite{pisoni2019interactive} and fostering heightened connections between remote visitors and artwork \cite{pisoni2020mediating}.



\subsection{VR Supports Museums Visitation for Older Adults}

With the proliferation of VR museums in the market, researchers have increasingly directed their efforts toward optimizing the design of these immersive spaces \cite{10.1145/3369394, 10.1145/3615522.3615544, 10.1145/3013971.3014018, 10.1145/3609987.3610006}. For instance, Diakoumako et al. investigated the efficacy of VR escape games as educational tools within archaeological museum settings \cite{10.1145/3609987.3610006}. However, these VR museum experiences have been tailored for general audiences, there exists a lack of research evaluating their effectiveness specifically among older adults.

Considerable research has underscored the distinct attributes characterizing the older demographic within museum contexts \cite{lafontaine2022virtual}.  Firstly, the purpose of older visitors to museums is not only to learn knowledge, but also to see museums as an important opportunity to socialize with fellow travelers \cite{hansen2014older, lafontaine2022virtual} or family members. Secondly, disparities in cognitive abilities and styles between older and younger cohorts have been observed \cite{kostoska2015virtual}. Previous research has found that more young people in certain cognitive styles, like ENTP, and older adults in other cognitive styles like INTJ \cite{antoniou2008reflections, antoniou2010modeling}. Additionally, older adults commonly encounter challenges in adapting to new technologies, experiencing disorientation and navigational difficulties within VR settings \cite{castilla2013process}. Thirdly, the interaction patterns exhibited by older individuals within museum environments are distinctive. For instance, Rainoldi et al. identified prolonged average fixation durations on exhibition information boards among older adults \cite{rainoldi2020museum}. However, given the impaired vision of older adults, it may be difficult for them to read the text on the information board. 

Until now, there exists a dearth of research investigating the challenges and preferences encountered by older demographics in engaging with VR museums. Our study stands as one of the pioneering endeavors delving into the VR museum encounters of this specific cohort. We not only conducted an preliminary inquiry into their challenges and preferences but also identified the potential design implications of our findings and outlined pertinent avenues for future research to better enhance the accessibility of VR museums for older individuals.

\section{METHOD}

We conducted a user-centered participatory workshop with 6 participants to better understand the challenges of and preferences for visiting VR museums. The workshop consists of three parts, as shown in~\autoref{process}: (i) background interview and introduction to VR technology and VR museum; (ii) VR experience session; (iii) semi-structured interview. We present the details of these steps in the following subsections. All interviews were audio recorded and lasted between one hour and an hour and a half.

\subsection{Participants}
We sourced our participants through a multifaceted approach, which included distributing informative posters with our contact information on social media platforms and actively engaging with local universities that cater to older adults. Our cohort of respondents boasts a diverse array of backgrounds. Table \ref{tab:table1} shows a summary of participants' information.
Participants ranged in age from 60 to 70 years old, comprising four females and two males. Two of them had experience of using VR. One participant had previously visited a VR museum before. All participants had visited museums physically.


\raggedbottom
\begin{table*}[htbp]
\centering
\caption {\label{tab:table1} Demographic Information of Participants}
\scalebox{1}{
\begin{tabular} {lccccc}
\toprule
\textbf{ID}  &\textbf{Age}   &\textbf{Gender}         &\textbf{VR experience}  &\textbf{VR museums experience} \\
\midrule
$P1$    & 70  & Male &yes &yes \\
$P2$    & 61  & Female &yes &no \\
$P3$    & 62  & Male &no &no \\
$P4$    & 60  & Female &no &no \\
$P5$    & 64  & Female &no &no \\
$P6$    & 60  & Female &no &no \\
\bottomrule
\end{tabular}
}
\label{table:01}
\Description{Demographic Information of Participants}
\end{table*}

\subsection{Apparatus and Settings}

\textbf{Devices.} We used Oculus Quest 2 as the apparatus. Participants interacted with the VR system using two hand controllers. Two researchers observed their physical actions, and we also projected the head-mounted display (HMD) projection onto the PC's screen.

\textbf{Platform.} We chose two VR museum applications on Steam VR, including \textit{Mona Lisa: Beyond the Glass} and \textit{Museum of Other Realities}, as our example platforms in the experiment. We selected these two VR museums because they possess the following key features: 1) They have diversified interactions, such as the ability to manipulate objects, the ability to listen to audio narration and watch video explanations; 2) They can switch multiple languages; 3) They offer multiplayer functionality and community engagement.

\begin{figure}[htbp]
\centering
\includegraphics[width=\textwidth]{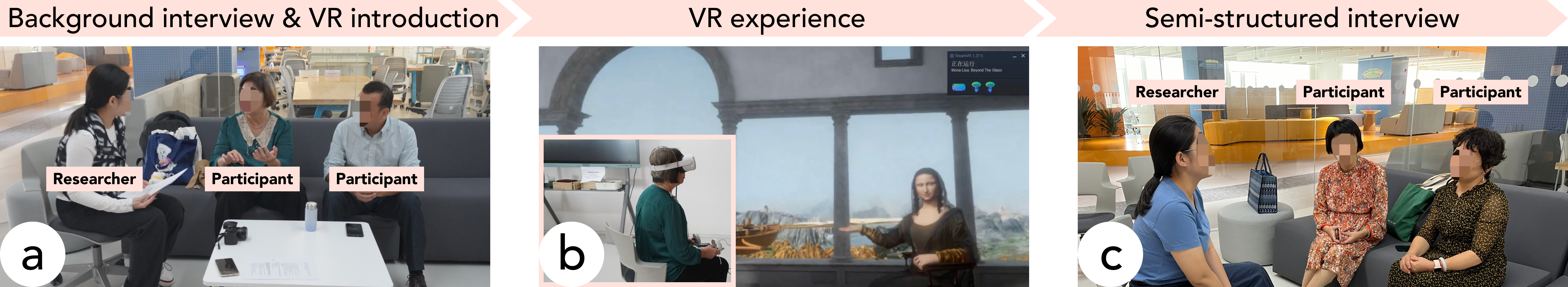}
\caption{\textbf{Participatory workshop process: a) A researcher conducted a background interview about participants' demographic information and introduced VR equipment to them; b) A participant was experiencing the VR museum; c) After the VR experience session, a researcher was interviewing two participants to understand their challenges, preferences, and expectations during VR museum visitation.}}
\label{process}
\end{figure}

\subsection{Procedure}

The participatory workshop consists of the following three parts (\autoref{process}):  

\textbf{Part One: Background Interview and VR Introduction.} This phase commenced with an overview of the session followed by a comprehensive background interview covering demographic aspects such as age, gender, prior museum visitation, familiarity with VR devices, and previous experiences with VR museum visits. Subsequently, participants were introduced to the VR technology and the VR museum, allowing them adequate time for acclimatization to the virtual environment.

\textbf{Part Two: VR Experience.} Under the guidance of experimenters, participants embarked on the VR experience. During the VR experience, participants had the flexibility to either stand or sit according to their personal preferences.

\textbf{Part Three: Semi-structured Interview.} This phase aimed to capture participants' challenges and preferences regarding their VR museum visitation. The interview encompassed a series of guided questions pertinent to the study objectives, including inquiries such as: (i) comparative assessments between VR museum experiences and prior physical museum visits; (ii) elucidation of encountered challenges and preferences during the VR museum tours; and (iii) suggestions for specific augmentations desired in VR museum experiences tailored for older audiences.

\subsection{Analysis Method}

Thematic analysis \cite{alhojailan2012thematic} was utilized to analyze the interview data in this study. Firstly, all recordings were transcribed verbatim into text. The research team then employed an open coding \cite{corbin1990grounded} by reading the transcripts to familiarize themselves with the data. Two coders independently coded the data, and the coding was discussed with the rest of the team during weekly research meetings. The team iteratively revised the coding during these meetings. An online tool called Miro was used to perform an affinity diagram analysis of the codes, which allowed the team to group the codes and identify common themes from the data. The findings were based on these themes, sub-themes, and codes.

\section{FINDINGS}


Our research preliminary elucidate the \textbf{challenges} and \textbf{preferences} among older adults during VR museum visitation.

\subsection{Challenges}

 \subsubsection{Challenges in spatial orientation and navigation}
Our participants encountered challenges in spatial orientation and navigation within the VR museum environment. A substantial portion of the participants (N=5) raised inquiries such as \textit{"Where am I?"} or \textit{"How do I get to this (specific location)?"} P5 noted encountering analogous challenges in physical museums, explicating: 
\begin{quote} 
\textit{"Museums often have large collections and varying themes from gallery to gallery. I often find myself spending a long time scrutinizing the signage before finding the artwork I want. The process that is both tiring and time-consuming."}
\end{quote}
P5 envisioned that VR museums hold the potential to provide a more simplified navigation approach. She proposed:
\begin{quote} 
\textit{"VR museums could integrate a feature facilitating the search for specific exhibits. For example, if I express an interest in a particular bronze artifact, I could utilize voice commands with the VR device, enabling it to guide me directly to the corresponding exhibition area."}
\end{quote}

\subsubsection{Challenges in understanding the explanations} Participants (N=4) conveyed discontentment with the interpretive descriptions provided by the system, asserting that the explanations provided by the system are often not tailored to older adults's level of knowledge. For instance, P1 opined: \textit{"The explanations are excessively specialized, and the information presented exceeds my level of knowledge, I find it somewhat challenging to apprehend."} They (N=2) expressed a preference for interpretive content to be tailored in accordance with the visitor's proficiency. As elucidated by P2:
\begin{quote} 
\textit{"I prefer to comprehend straightforward and accessible explanations as I am relatively new to this subject. However, other people with a foundational understanding may seek more specialized content. Providing a singular version of interpretive descriptions is inadequate."}
\end{quote}
Furthermore, participants (N=2) expressed a desire for the system to address queries from visitors. P1 emphasized: \textit{"If comprehension remains elusive after listening to the explanation, I hope to pose questions, this would enhance my understanding of the exhibits."}

\subsection{Preferences}

\subsubsection{Aspire heightened multi-sensory interactions} Our participants actively partook in interactive engagements with the exhibits during the VR experience, as shown in Figure 2. Every participant conveyed a strong affinity for these interactive elements, with a subset of participants (N=4) expressing a desire for more immersive interactive encounters. For example, P2 expressed an aspiration to \textit{"try on"} and \textit{"capture photographs while adorned in"} exquisite handicrafts. P3 articulated a wish to personally disassemble mechanical artifacts, elucidating: \textit{"Beyond the surface, a television is merely a flat screen. The intricacies of how it emanates light and generates images through diverse components are captivating."} Additionally, participants (N=3) voiced a preference for these interactions to be multi-sensory, P2 exemplified this preference, stating: \textit{"Upon opening a perfume, I would like to discern its fragrance."}

\begin{figure}[htbp]
\centering
\includegraphics[width=\textwidth]{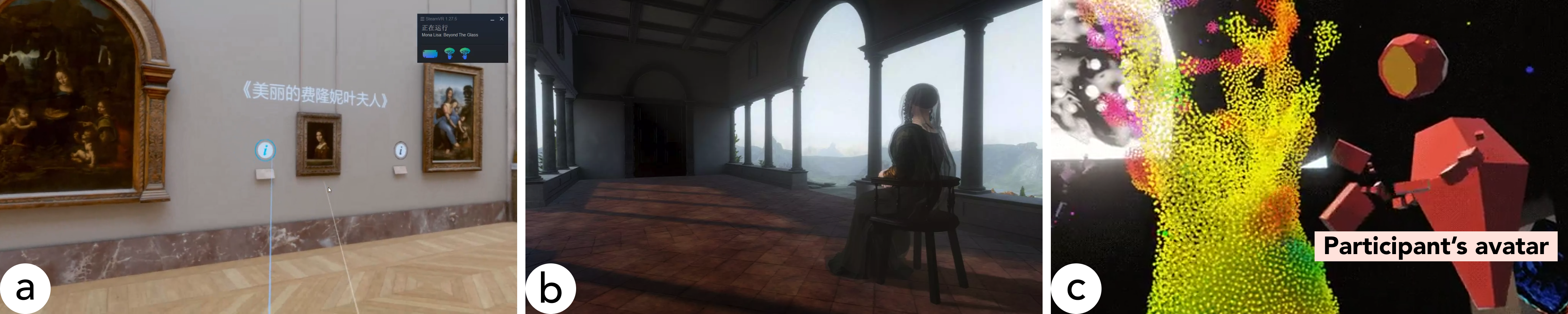}
\caption{\textbf{Participants interacting with the exhibits during the VR experience: a) Clicking the button to get audio narration; b) Getting into the VR scene of the world-famous painting Mona Lisa; c) Touching the particle effect of exhibits through their avatars.}}
\label{Figure}
\end{figure}

\subsubsection{Hope to visualize the imagined content in their mind} When visiting VR museums, our participants wish to visualize the scenes or objects they imagine to  facilitate communication with others. The majority of participants (N=5) regard VR museum visits as exceptional opportunities for social engagement. For instance, P4 conveyed an intention to engage in future VR museum visits with \textit{"family members, former colleagues, or new acquaintances,"} recognizing these visits as platforms for shared discussion. P2 elucidated:
\begin{quote} 
\textit{"If the artifacts I encounter have relevance to my past experiences or prior knowledge, such as urban landscapes from the 80s or artifacts seen in another museum, I tend to form associations and wish to share these insights with others."}
\end{quote}
To enhance communication, participants (N=4) expressed a desire to reconstruct the mental associations they associated regarding scenes or objects, thereby facilitating a more vivid and intuitive sharing experience.

\subsubsection{Desire different exploration modes} Our participants manifested diverse preferences regarding exploration modes. While the majority of participants (N=5) regard VR museum visits as exceptional opportunities for social engagement, P1 conveyed a predilection for solitary exhibition viewing, opting to subsequently engage in discussions with family and friends.
P1 rarely engaged in conversation during museum visits. This was because he perceived it as impolite to initiate discussions when others were deeply immersed in their observations. Additionally, the limited time immediately after exploring the museum constrains the opportunity for in-depth conversations.
As he highlighted: \textit{"Post-exit discussions are typically brief due to visitors having just concluded their viewing, and many are hastening to the subsequent event, such as partaking in a meal."} Consequently, he favored \textit{"selecting a time when all participants are available to specifically deliberate on the museum visiting experience."}

Considering the variegated preferences among participants concerning exploration modes, VR museum visits could incorporate features enabling visitors to tailor their experiences. As expressed by P4: \textit{"It should accommodate the selection of modes, allowing one person to relish the experience alone, or alternatively, facilitating joint participation for two people."}

\section{DISCUSSION}
Through an interview-based investigation, we preliminarily found the challenges and preferences of older adults during VR museum visits.
In this section, we further discuss the potential design implications of our preliminary findings and outline pertinent avenues for future research to better enhance the accessibility of VR museums for older adults.

\subsection{Designing Accessible Approaches for Navigation in VR Museums}

Although virtual museums aim to provide "equitable opportunities for mobility-challenged older adults to visit a museum" \cite{hilton2019don}, our study reveals that older adults still face challenges with location and navigation in VR museums. These difficulties may stem from complex VR interfaces, sensory overload, unfamiliarity with digital interactions, and so on.
To address this issue and make VR museums more intuitive, future work could explore more user-friendly interaction methods. Firstly, one promising solution is the implementation of voice transportation \cite{hombeck2023tell}. By using simple voice commands, visitors can effortlessly explore the virtual space, bypassing traditional navigation complexities. Secondly, gesture-based systems \cite{gestureVR} could offer an intuitive alternative. Through straightforward hand movements, visitors know exactly which direction to go. Thirdly, tactile guidance feedback \cite{Cugnet2017AVW}, which provides users with vibrotactile information, can offer a form of haptic navigation. This method helps in compensating for the lack of physical cues in a virtual space, guiding users through vibrations that indicate direction and points of interest.
By adopting these approaches, VR museums ensure inclusivity and enhance the navigation experience for all visitors, particularly for older adults who may not be as familiar with digital technology.


\subsection{Providing Personalized Interpretations to Enhance Visitors' Understanding of Exhibits}

Cognitive abilities tend to decline gradually among older adults \cite{antoniou2008reflections, kostoska2015virtual}. Our research indicates that the interpretive descriptions provided by the system are often not tailored to older adults's level of knowledge. In order to address this challenge, VR museums could offer interpretive content tailored to visitors' knowledge and cognitive levels, while also catering to their specific inquiries. AIGC (Artificial Intelligence Generated Content) is an innovative AI technology that encompasses tasks such as generating images from text, producing text from text, and translating text from images \cite{leng2023codp}. Past studies have utilized AIGC tools to assist children in language acquisition \cite{leng2023codp}. Future research could explore how AIGC tools can deliver personalized interpretations for older adults in VR museums. This may involve training the AIGC tool to recognize the language patterns of older adults, assess their cognitive levels, and customize interpretive content to be more simplified or in-depth based on their comprehension and expertise. By providing personalized interpretations, VR museums can significantly enhance the engagement and understanding of older adults during visiting experiences.

\subsection{Designing Multi-sensory Interactive Experiences}

Previous research indicates that older adults tend to prefer reading text over engaging with exhibits in traditional museums \cite{rainoldi2020museum}. In contrast, our study reveals that in VR museums, older adults share the same enthusiasm for rich interactive experiences as their younger counterparts and anticipate that these interactions will offer diverse sensory feedback, enhancing their sense of realism. Researchers have explored various sensory feedback modalities, including tactile feedback \cite{10.1145/3027063.3050426, 10.1145/3411764.3445285}, olfactory feedback \cite{10.1145/3544549.3585636, 10.1145/3615522.3615544}, and gustatory feedback \cite{10.1145/3411763.3441343}. Future research should delve into integrating these sensory interactions into the VR museum experience and investigate their impact on older adults' museum experience, such as enhancing their engagement or aiding a better understanding of artworks.

\subsection{Visualizing the Imagined Content in Older Visitors' Mind}

Our research shows that older adults view visiting VR museums as excellent social opportunities and wish to visualize the content they envision to facilitate communication with other visitors. Recent studies have begun to explore how AIGC tools can assist people in expressing ideas from their minds and generating corresponding images \cite{10.1145/3544548.3580999}. However, these applications primarily focus on professional contexts, such as generating product drawings based on designers' textual descriptions, aiming to improve work efficiency \cite{10.1145/3544548.3580999}. There is currently a lack of research investigating how AIGC tools can help older adults visualize what they imagine and thus foster communication. Two points should be noted: firstly, each person's memory is unique, which makes it difficult for AIGC tools to accurately reproduce identical scenes or objects. Secondly, individual memory may become blurry over time, and this lack of clarity adds a nostalgic aesthetic to recollections. Future work could consider recreating the atmosphere of fuzzy scenes, emphasizing certain elements in the scenes (such as decorating styles, specific objects, etc.). Experimental approaches may further clarify how to reproduce the atmosphere of scenes and which elements should be emphasized.

\subsection{Providing Both Individual and Social Exploration Modes}

Building on prior research that highlights the distinct experiences of single-player versus multiplayer modes in VR environments \cite{hansen2020asymmetrical}, our participants showed diverse preferences in terms of interaction with other visitors. Future work should consider offering various personalized interaction modes in order to better meet the diverse preferences of participants when interacting with other visitors. On one hand, individual exploration modes should be provided for solo visitors, allowing them to delve into the exhibits according to their own paces and interests. 
On the other hand, for those who enjoy sharing experiences and interacting with other visitors, a social interaction mode should be provided. In this mode, visitors can communicate with others in real-time via voice chat or instant messaging, sharing their insights about the exhibition. Additionally, a virtual "social corner" could be designated to serve as a gathering place for users to participate in thematic discussions, thus fostering deeper social connections.

\section{LIMITATION AND FURURE WORK}





One limitation was that we had a limited number of participants. We conducted workshop with six participants to comprehend the challenges and preferences associated with visiting VR museums for older individuals. However, qualitative research methods involving a small participant pool may present limitations. Future work could benefit from a larger participant sample and the incorporation of quantitative methods, such as measuring the time participants spend searching for specific exhibits. Additionally, another limitation was that our research exclusively employed an existing virtual reality museum. In subsequent studies, designing a museum tailored for older adults based on our design thinking principles and inviting older individuals to visit could enhance our understanding of the effectiveness of our design approach.


\section{CONCLUSION}

Through the participatory workshop, our research sought to gain a preliminary understanding of the challenges and preferences faced by older adults during VR museum visits. The findings indicate that older participants universally encountered difficulties in spatial positioning and navigation within the VR museum environment. Moreover, they expressed a collective opinion that the interpretive descriptions provided by the system were not tailored to their cognitive capacities. There was a shared desire among participants for enhanced multi-sensory interactions, as well as a keen interest in visualizing the scenes or objects they imagined to facilitate more vivid and intuitive sharing experiences. Additionally, participants exhibited diverse preferences in interacting with other visitors. This study has formulated a set of potential design guidelines aimed at improving the accessibility of VR museums for older adults.

\bibliographystyle{ACM-Reference-Format}
\bibliography{references}

\appendix

\end{document}